\documentclass[proof]{WileyASNA-v1}

\articletype{Article Type}%

\received{26 July 2020}
\revised{26 July 2020}
\accepted{26 July 2020}

\usepackage{xcolor}
\begin{document}

\title{A Study of Open Clusters Frolov 1 and NGC 7510 using CCD UBV Photometry and Gaia DR2 Astrometry}

\author[1]{Talar Yontan*}

\author[1]{Sel\c cuk Bilir}

\author[1]{Tansel Ak}

\author[2]{Burcu Akbulut}

\author[2]{Remziye Canbay}

\author[3,4]{Timothy Banks}

\author[5]{Ernst Paunzen}

\author[1]{Serap Ak}

\author[1]{Z. Funda Bostanc\i}

\authormark{Yontan \textsc{et al.}}

\address[1]{Istanbul University, Faculty of Science, Department of Astronomy and Space Sciences, 34119, University, Istanbul, Turkey}

\address[2]{Istanbul University, Institute of Graduate  Studies in Science, Programme of Astronomy and Space Sciences, 34116, Beyaz{\i}t, Istanbul, Turkey}

\address[3]{Nielsen, Data Science, 200 W Jackson Blvd \#17, Chicago, IL 60606, USA}

{\color{blue} \address[4]{Physics \& Astronomy, Harper College, 1200 W Algonquin Rd, Palatine, IL 60067, USA}}

\address[5]{Department of Theoretical Physics and Astrophysics, Masaryk University, Kotl\'a\u rsk\'a 2, 611 37 Brno, Czech Republic}

\corres{*Talar Yontan, Istanbul University, Faculty of Science, Department of Astronomy and Space Sciences, 34119, University, Istanbul, Turkey. \email{talar.yontan@istanbul.edu.tr}}

\abstract{We present reddening, photometric metallicity, age and distance estimates for the Frolov~1 and NGC 7510 open clusters based on CCD {\it UBV} photometric and {\it Gaia} data. Photometric observations were collected using the 1-m telescope of the T\"UB\.ITAK National Observatory. {\it Gaia} DR2 proper motion data in the direction of two groupings were used to identify cluster membership. We determined mean proper motion values ($ \mu_{\alpha}\cos\delta, \mu_{\delta}$) = ($-3.02\pm 0.10$, $-1.75 \pm 0.08$) and ($-3.66 \pm 0.07$, $ -2.17 \pm 0.06$) mas yr$^{-1}$ for Frolov 1 and NGC 7510, respectively. We used two-colour diagrams to obtain $E(B-V)$ colour excesses for Frolov~1 and NGC 7510 as $0.65\pm0.06$ and $1.05\pm0.05$ mag, respectively. We derived the photometric metallicity of Frolov 1 as [Fe/H] = 0.03$\pm$0.03 dex and adopted a solar metallicity for NGC 7510. Based on these reddening and metallicities we determined the distance moduli and ages of the clusters via fitting PARSEC isochrones to the cluster colour-magnitude diagrams. Isochrone fitting distances for Frolov 1 and NGC 7510 are $2,864 \pm 254$ and $2,818 \pm 247$ pc, respectively, which correspond to the ages $35 \pm 10$ Myr and $18 \pm 6$ Myr. We also calculated mean {\it Gaia} distances and compared them with those given in the literature and in this study, concluding that our results are in good agreement with previous work. Finally, we calculated the mass function slopes as being $X=-1.21\pm0.18$ for Frolov 1 and $X=-1.42\pm0.27$ for NGC 7510.}


\jnlcitation{\cname{%
\author{T. Yontan}, 
\author{S. Bilir}, 
\author{T. Ak}, 
\author{B. Akbulut}, 
\author{R. Canbay},
\author{T. Banks},
\author{E. Paunzen},
\author{S. Ak},
and
\author{Z. Funda Bostanc\i}
} (\cyear{XXXX}), 
\ctitle{A Study of Open Clusters Frolov 1 and NGC 7510 using CCD UBV Photometry and Gaia DR2 Astrometry}, \cjournal{AN}, \cvol{XXX;XX:X--X}.}
\maketitle

\section{Introduction}\label{sec1}
Open clusters are valuable tools in understanding numerous features of the Galaxy and stellar evolution. As such a grouping is formed from the collapse of a molecular cloud, the metallicities, distances and ages of a cluster's stars are similar although masses differ from star to star. The study of open clusters with different individual ages gives insight into stellar properties and evolution stages. In addition to this, inter-relations between the accurately determined astrophysical parameters (such as distance-metallicity and age-metallicity) of numerous open clusters enable us to understand the chemical and temporal evolution of their surroundings, such as the Galactic disc. Open clusters are gravitationally weakly bound systems of limited volumes and stellar numbers (compared to globular clusters, for example), with their component stars sharing similar spatial velocities. The main astrophysical parameters (colour excess, metallicity, distance, and age) of a cluster can be obtained from the colour-magnitude (CMD) and two-colour (TCD) diagrams  through comparison with the results of stellar models, such as isochrones. Open clusters have different morphologies \citep{Trumpler30}: some clusters are more prominent than  surrounding field stars because of their relative stellar over-densities, while it can be difficult to distinguish the sparse clusters from the field stars. These properties affect the observed characteristics and distributions in the CMDs and TCDs of the open clusters such as main sequence, turn off, and the red clump regions. To determine precise astrophysical parameters, not only are features in the CMDs and TCDs important but also the accurate selection of cluster member stars, together with the use of homogeneous data and methods during analysis. Studies by different authors of the same open cluster can have quite different values for astrophysical parameters \citep{Dias02, Netopil16}. The quality of data combined with outdated isochrones, cluster membership determination techniques, and analysis methods can affect the results and cause degeneracy between determined parameters of clusters \citep{Anders04, King05}.

The {\it Gaia} data release 2 \citep[{\it Gaia} DR2,][]{Gaia18} provides unprecedented astronomical information for more than 1.3 billion objects. These data contain positions ($\alpha$, $\delta$), parallaxes ($\varpi$) and proper motions ($\mu_{\alpha}\cos\delta, \mu_{\delta}$) with a limiting magnitude of $G\sim 21$ mag. For $G\leq 15$ mag and $G\sim 17$ mag, the uncertainties of parallaxes are up to 0.04 mas and 0.1 mas respectively. Uncertainties of the proper motion components are up to 0.06 mas yr$^{-1}$ in the magnitude range up to $G\leq 15$ mag, dropping to 0.2 mas yr$^{-1}$ by $G\sim 17$ mag. Such precise data allow us to sensitively select physical cluster members as well as to calculate mean proper motion components and {\it Gaia} distance values for these open clusters \citep[c.f.][]{Cantat-Gaudin18, Cantat-Gaudin20}.

\begin{table*}[htbp]
\setlength{\tabcolsep}{4pt}
  \centering
  \caption{Basic parameters determined for Frolov 1 and NGC 7510 and compiled from literature studies. Columns denote cluster name, colour excess ($E(B-V)$), distance modulus ($\mu_V$), distance ($d$), iron abundance ([Fe/H]), age ($t$), mean {\it Gaia} distance ($d_{Gaia}$) and mean proper motion components ($\mu_{\alpha} \cos \delta, \mu_{\delta}$) for each cluster.}
    \begin{tabular}{cccccccccc}
    \hline
    \hline
Cluster & $E(B-V)$ & $\mu_V$ & $d$ & [Fe/H] & $t$ & $d_{Gaia}$ & $\mu_{\alpha} \cos \delta $ & $\mu_{\delta}$ & Reference \\
  & (mag)  & (mag)  & (pc) & (dex) & (Myr) & (pc) & (mas yr$^{-1}$) & (mas yr$^{-1}$)\\
\hline
\hline
Frolov 1 &  0.60 & 13.90 & 2,560 & --- & 46 & --- & -5.91$\pm$0.65 & -1.35$\pm$1.22 & (1)\\
         &  0.65$\pm$0.06 & 14.30$\pm$0.20 & 2,864$\pm$254 & 0.03$\pm$0.03 & 35$\pm$10 & 2,800$\pm$136 & -3.02$\pm$0.10 & -1.75$\pm$0.08 & (2)\\
\hline
NGC 7510 & 0.90   & 14.38 & 2,075 & --- & 16 & --- & -3.63$\pm$0.89 & -0.92$\pm$2.00 & (1) \\
      &   1.05$\pm$0.05 & 15.50$\pm$0.20 & 2,818$\pm$247 & 0 & 18$\pm$6 &  3,450$\pm$477 & -3.66$\pm$0.07 & -2.17$\pm$0.06 & (2) \\
      &   0.86 & 14.25 & 2,075 & --- & 38 & --- & --- & --- & (3) \\
      &   1.16  & 16.07 & 3,125 & --- & ---& --- & --- & --- & (4) \\
      &   1.00 & 14.40$\pm0.30$ & 3,160$\pm$450 & --- & 10 &  --- & --- & --- & (5) \\
      &   1.12  & 15.92 & 3,090 & --- & 10 &  --- & --- & --- &  (6) \\
      &   0.90   & 15.50$\pm0.25$ & 3,480$\pm$420 & --- & 22 & --- & --- & --- & (7) \\
\hline
\hline
    \end{tabular}%
\\
    (1) \citet{Kharchenko05}, (2) This study, (3) \citet{Hoag61}, (4) \citet{Fenkart85}, (5) \citet{Sagar91}, (6) \citet{Barbon96}, (7) \citet{Paunzen05}
  \label{tab:addlabel}%
\end{table*}%

In this study, the main goal was to derive structural, astrophysical and astrometric parameters of two poorly studied clusters in the northern hemisphere based on wide-field CCD {\it UBV} photometry.  We analysed {\it Gaia} DR2 astrometric data ($\varpi, \mu_{\alpha}\cos\delta, \mu_{\delta}$) in conjunction with these images, utilising independent methods to investigate mean proper motion, {\it Gaia} distance, main astrophysical parameters (reddening, metallicity, distance and age) and the mass functions for these two clusters: Frolov 1 and NGC 7510. The calculated physical parameters for the clusters are given in Table 1. Information about previous studies of two clusters is given below.

\subsection{Frolov 1}
\citet{Frolov77} presented photographic photometry of four open clusters in the constellation Cassiopeia, including Frolov 1 ($\alpha=23^{\rm h} 57^{\rm m} 25^{\rm s}, \delta=+61^{\rm o} 37^{'} 48^{''}$, $l=116^{\rm o}.56$, $b=-0^{ \rm o}.57$) which he noted is composed of a low density of faint stars. No deep CCD observations or similar studies are available in the literature for Frolov 1. Kharchenko et al. (2005, 2013) give a general analysis of the stars located in the cluster, as detected from the ASCC-2.5 \citep{Kharchenko01} and Tycho-2  catalogues \citep{Hog00}. They determined the distance and age of Frolov 1 as 2,560 pc and $\log t = 7.66$ yr, respectively.

\subsection{NGC 7510}
This cluster is located in Cepheus (with coordinates $\alpha=23^{\rm h} 11^{\rm m} 03^{\rm s}, \delta=+60^{\rm o} 34^{'} 12^{''}$, $l=110^{\rm o}.90$, $b=+0^{ \rm o}.06$) and classified as Trumpler II3r \citep{Lynga87}. There is substantial variation in literature estimates for the distance of NGC 7510. While \citet{Trumpler30} determined the distance of the cluster as 5,040 pc, \citet{Buscombe53} calculated this value as 2,200 pc. \citet{Johnson61} indicated that this difference in estimated cluster distance is caused by an inhomogeneous interstellar medium in the direction of the cluster. \citet{Hoag61} studied 91 stars within $V=16$ mag in the {\it UBV} system with photoelectric and photometric observational techniques. They determined the reddening, distance, and age of the cluster  via their CMD as $E(B-V)=0.86$ mag, $d=2,075$ pc, and $t=38$ Myr, respectively. A detailed study of {\it UBV} photographic photometry of NGC 7510 was made by \citet{Fenkart85}. They analysed 314 stars located through the cluster region within $10<V<18$ mag using $V\times B-V$ and $V\times U-B$ CMDs and calculated the $\mu_V$ distance modulus, $E(U-B)$, and $E(B-V)$ reddenings as 16.07, 0.88, and 1.16 mag, respectively. \citet{Sagar91} utilized CCD {\it BVI} photometric observations of the cluster. They analysed 592 cluster stars down to a limiting $V$ magnitude of 21, showing that the reddening of NGC 7510 is inhomogeneous and changes between $1<E(B-V)<1.3$ mag. In the study, the distance moduli was determined as $(\mu_V)_0=12.50\pm 0.30$ mag and the age of the cluster was calculated from the CMDs as $t=10$ Myr. NGC 7510 was also studied by \citet{Barbon96}. In the study, photographic {\it UBV} observations of stellar fields located in the cluster region were collected using the 182 and 122 cm telescopes at Asiago. Stellar magnitudes were measured with iris photometry from a total of 20 photographic plates consisting of eight, six and six exposures in the $U$, $B$, and $V$ filters, respectively. \citet{Barbon96} standardised these magnitudes via the {\it UBV} photoelectric observations of \citet{Hoag61}. They constructed a $U-B\times B-V$ TCD of the cluster stars. Fitting the ZAMS of \citet{Iriarte58} on this diagram, the colour excesses of the NGC 7510 were determined as $E(U-B)=0.87$ and $E(B-V)=1.12$ mag. The de-reddened distance modulus of the cluster was calculated via main-sequence matching on the $V\times B-V$ CMD as ($\mu_V)_0=12.45$ mag. Using the same diagram and \citet{Bertelli94}'s isochrones, \citet{Barbon96} determined the age of the cluster as 10 Myr. To conclude the literature review, we note that \citet{Paunzen05} analysed the peculiar stars in NGC 7510 with CCD photometry and investigated the reddening, distance modulus, distance and age of the cluster finding $E(B-V)=0.90$ mag, $\mu_V=15.50\pm0.25$ mag, $d=3,480\pm420$ pc, and $t=22$ Myr respectively.

\section{Observations}\label{sec2}
The observations of the two open clusters were performed using a Fairchild 486 BI 4k$\times$4k CCD camera with Johnson-Cousins {\it UBV} filters attached to the $f/8$ Cassegrain focus of the 1-m Ritchey Chretien telescope (T100) of the T\"UB\.ITAK National Observatory (TUG) located in Antalya, Turkey. The pixel scale of the telescope-camera system is $0''.31$ pixel$^{-1}$, which gives about a $21^{'} \times 21^{'}$ field of view on the sky. The gain and readout noise of the CCD are 0.57 electrons ADU$^{-1}$ and 4.11 electrons, respectively. We used long and short observation exposure times to obtain a range of limiting and saturation magnitudes for the $U$, $B$ and $V$ filters, allowing good detection of both brighter and fainter stars located in the cluster regions. Inverse coloured $V$ band images of the two clusters are displayed in Fig. 1. The short and long exposure frames were separately combined to form two longer duration images for analysis, for each filter and cluster. The observational log is given in Table 2. Photometric calibrations were based on \citet{Landolt09} standard stars. A total of 88 stars in 13 selected areas were observed. Details are given in Table 3. The air masses ($X$) were between 1.2 and 1.9 for the standard stars monitored during the observation nights. For the photometric calibrations, we observed standard stars of \citet{Landolt09} in {\it UBV} bands during the observing run. Standard CCD reduction techniques were applied using IRAF\footnote{IRAF is distributed by the National Optical Astronomy Observatories} packages for the pre-processing of all the images which were taken. Astrometric corrections were applied using PyRAF\footnote{PyRAF is a product of the Space Telescope Science Institute, which is operated by AURA for NASA} and astrometry.net\footnote{http://astrometry.net} routines together with our own scripts for the cluster images. Using IRAF's aperture photometry packages, we measured the instrumental magnitudes of \citet{Landolt09} stars. Then we applied multiple linear fits to these magnitudes and derived photometric extinction and transformation coefficients for each observing night (Table 4). Source Extractor (SExTractor) and PSF Extractor (PSFEx) routines \citep{Bertin96} were applied to measure the instrumental magnitudes of the objects located in the cluster fields. We applied aperture corrections to these magnitudes and transformed them to standard magnitudes in the Johnson photometric system using transformation equations as described by \citet{Janes13}. 

\begin{figure}
\centering
\includegraphics[scale=.27, angle=0]{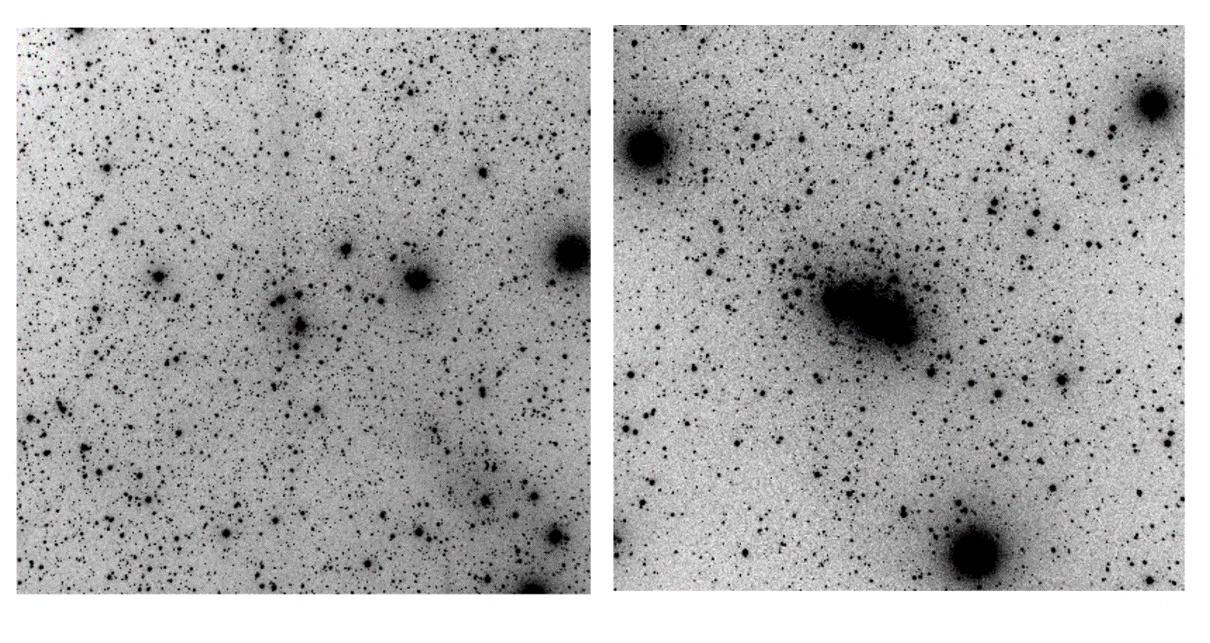}
\caption{Negative long exposure time $V$-band images of the two clusters:  the total exposure times are 1200 seconds for Frolov 1 (left panel) and 150 seconds for NGC 7510 (right panel). The image dimensions of a square field of view are $21^{'} \times 21^{'}$. East is to the left in these diagrams, and North is upwards.} 
\end {figure}

\begin{figure*}
\centering
\includegraphics[scale=.55, angle=0]{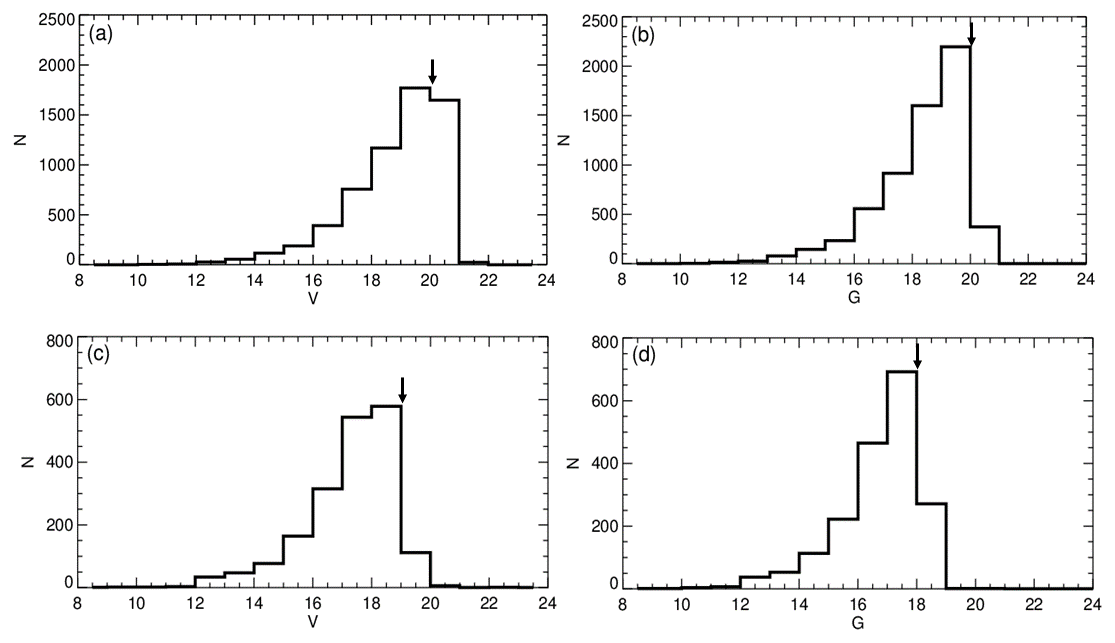}\\
\caption{Interval $V$ and $G$-band magnitude histograms of Frolov 1 (a, b) and NGC 7510 (c, d). Arrows indicate the completeness limits in $V$ and $G$ magnitudes.} 
\end {figure*}

\begin{table}
\setlength{\tabcolsep}{5pt}
  \centering
  \caption{Log of observations. Columns represent cluster name, observation date, filters, exposure times and number of exposure ($N$).}
    \begin{tabular}{lcccc}
\hline
 Cluster  &  Observation& \multicolumn{3}{c}{Filter~/~(Exposure Time (s)$\times~N$)} \\
          & Date & $U$ & $B$ & $V$ \\
\hline
Frolov 1  & 02.11.2016 &   25$\times$3  & 2$\times$3    & 1.2$\times$4\\
          &            &  1200$\times$2 & 1200$\times$2 & 600$\times$2\\
NGC 7510  & 06.11.2018 &   60$\times$4  & 6$\times$5    & 3$\times$5\\
          &            &  1800$\times$1 & 60$\times$5   & 30$\times$5\\
\hline
    \end{tabular}
\end{table}

\begin{table}[htbp]
  \centering
  \caption{Information on the observations of standard stars from selected \citet{Landolt09} fields. The columns are the observation date, star field name as from Landolt, the number of standard stars ($N_{\rm st}$) observed in a given field, the number of pointings to each field ($N_{\rm obs}$, i.e., observations), and the airmass range the fields (on a given night) were observed over ($X$).}
    \begin{tabular}{ccccc}
    \hline
Date	  & Star Field	& $N_{\rm st}$	& $N_{\rm obs}$	& $X$\\
\hline
          & SA92    & 6         & 1	        &          \\
	      & SA93	& 4	        & 1	        &          \\
	      & SA95	& 9 	    & 2	        &          \\
          & SA96	& 2	        & 3	        &          \\
          & SA97	& 2	        & 1	       &          \\
02.11.2016 & SA98	& 19	    & 1	        & 1.22 - 1.88\\
	      & SA110	& 10	    & 1	        &          \\
	      & SA111	& 5	        & 1	        &          \\
	      & SA112	& 6	        & 2	        &          \\
	      & SA113	& 15	    & 1	        &          \\
	      & SA114	& 5	        & 1	        &          \\
\hline	      
      	  & SA92    & 6         & 1	        &          \\
      	  & SA93	& 4	        & 1	        &          \\
      	  & SA94	& 2	        & 1	        &          \\
      	  & SA95	& 9 	    & 1	        &          \\
      	  & SA96	& 2	        & 2	        &          \\
06.11.2018 & SA97    & 2	        & 2	        &  1.23 - 1.87 \\
      	  & SA98	& 19	    & 1	        &          \\
          & SA99	& 3	        & 1	        &          \\
	      & SA110	& 10	    & 1         &          \\
	      & SA111	& 5	        & 1	        &          \\
	      & SA112	& 6	        & 1	        &          \\
          & SA114	& 5	        & 1	        &          \\
          
    \hline
    \end{tabular}%
  \label{tab:addlabel}%
\end{table}%

\begin{table*}[htbp]
  \centering
  \caption{Derived transformation and extinction coefficients. $k$ and $k^{'}$ are primary and secondary extinction coefficients, respectively, while $\alpha$ and $C$ are transformation coefficients.}
    \begin{tabular}{cccccc}
 \hline
    Filter/Colour index & Observation Date & $k$ & $k'$ & $\alpha$ & $C$ \\
\hline
    $U$ & 02.11.2016 & 0.362$\pm$0.085 & -0.319$\pm$0.111 & -- & -- \\
    $B$ &            & 0.247$\pm$0.043 & -0.068$\pm$0.043 & 1.013$\pm$0.062 & 1.265$\pm$0.061 \\
    $V$ &            & 0.114$\pm$0.002 & -- & -- & -- \\
    $U-B$            & & -- & -- & 1.304$\pm$0.163 & 3.729$\pm$0.122 \\
    $B-V$            & & -- & -- & 0.077$\pm$0.011 & 1.377$\pm$0.030 \\
\hline
    $U$ & 06.11.2018 & 0.521$\pm$0.062 & -0.122$\pm$0.069 & -- & -- \\
    $B$ &            & 0.227$\pm$0.046 & -0.021$\pm$0.052 & 0.932$\pm$0.077 & 1.512$\pm$0.068 \\
    $V$ &            & 0.124$\pm$0.018 & -- & -- & -- \\
    $U-B$&           & -- & -- & 0.986$\pm$0.102 & 3.718$\pm$0.093 \\
    $B-V$&           & -- & -- & 0.073$\pm$0.007 & 1.570$\pm$0.028 \\
\hline
    \end{tabular}%
  \label{tab:addlabel}%
\end{table*}%

\begin{table*}[htbp]
  \centering
  \caption{Mean internal photometric errors of {\it UBV} and {\it Gaia} passbands, $V$, $U-B$, $B-V$, $G$ and $G_{BP}-G_{RP}$ against the $V$ magnitude. $N$ is the number of stars according to $V$ magnitude ranges.}
\begin{tabular}{c|cccccc|ccccccc}
\hline
 & \multicolumn{6}{c}{Frolov 1} & \multicolumn{6}{c}{NGC 7510}\\
\cline{2-13}
$V$ & $N$ & $\sigma_V$ & $\sigma_{U-B}$ & $\sigma_{B-V}$ & $\sigma_G$ & $\sigma_{G_{BP}-G_{RP}}$ & $N$ & $\sigma_V$ & $\sigma_{U-B}$ & $\sigma_{B-V}$ & $\sigma_G$ & $\sigma_{G_{BP}-G_{RP}}$\\
\hline
    ( 8, 12] &   11 & 0.001 & 0.005 & 0.001 & 0.001 &  0.002 &   8 & 0.001 & 0.002 & 0.002 & 0.001 &  0.002\\
    (12, 14] &   84 & 0.002 & 0.010 & 0.002 & 0.001 &  0.002 &  81 & 0.003 & 0.005 & 0.004 & 0.001 &  0.003\\
    (14, 15] &  115 & 0.003 & 0.010 & 0.003 & 0.001 &  0.003 &  77 & 0.003 & 0.007 & 0.003 & 0.001 &  0.003\\
    (15, 16] &  188 & 0.001 & 0.007 & 0.003 & 0.001 &  0.004 & 164 & 0.004 & 0.012 & 0.006 & 0.001 &  0.004\\
    (16, 17] &  394 & 0.002 & 0.013 & 0.003 & 0.001 &  0.006 & 315 & 0.008 & 0.026 & 0.011 & 0.001 &  0.006\\
    (17, 18] &  754 & 0.004 & 0.026 & 0.006 & 0.001 &  0.012 & 544 & 0.017 & 0.047 & 0.023 & 0.001 &  0.013\\
    (18, 19] & 1169 & 0.009 & 0.053 & 0.013 & 0.002 &  0.025 & 577 & 0.036 & 0.082 & 0.053 & 0.001 &  0.023\\
    (19, 20] & 1775 & 0.019 & 0.078 & 0.031 & 0.003 &  0.055 & 112 & 0.071 & 0.103 & 0.129 & 0.002 &  0.035\\
    (20, 22] & 1665 & 0.040 & 0.125 & 0.067 & 0.005 &  0.099 &   7 & --- & ---   & --- & --- &  ---\\
\hline
    \end{tabular}%
  \label{tab:addlabel}%
\end{table*}%

\section{Data Analysis}
\subsection{Photometric data}
Photometric catalogues listing all the detected stars in the cluster regions are available electronically for Frolov 1 and NGC 7510. We identified 6,155 sources for Frolov 1 and 1,885 sources for NGC 7510, before constructing their photometric and astrometric catalogues. Both of the catalogues contain positions ($\alpha, \delta$), apparent $V$ magnitudes, $U-B$ \& $B-V$ colours, proper motion components ($\mu_{\alpha}\cos\delta, \mu_{\delta}$) along with trigonometric parallaxes ($\varpi$) from the {\it Gaia} DR2, and membership probabilities ($P$) as calculated in this study. Magnitude and colour inaccuracies of Johnson ($V$, $U-B$, $B-V$) and {\it Gaia} DR2 ($G$, $G_{BP}-G_{RP}$) photometries are adopted as internal errors. We calculated mean photometric errors as a function of $V$ magnitude intervals (see Table 5). It can be seen from Table 5 that the mean internal errors in the {\it UBV} data for stars brighter than $V=22$ are smaller than 0.05 mag for Frolov 1 and 0.07 mag for stars brighter than $V=20$ mag in NGC 7510. Moreover, mean internal errors in {\it Gaia} DR2 data for the stars brighter than $V=22$ mag are smaller than 0.005 and 0.002 mag for Frolov 1 and NGC 7510, respectively.

In order to determine precise astrophysical parameters, we quantified the quality of data by taking into account the photometric completeness limit. The exposure duration used during observations creates a limiting magnitude where counting completeness drops away from unity. The clusters studied were not strongly affected by blending of the point spread functions. To find this limit, we constructed $V$ and $G$ magnitude  histograms of the two clusters. As can be seen from Fig. 2, we concluded that stellar counts reduce for magnitudes fainter than $V=20$ for Frolov 1 and $V=19$ mag for NGC 7510, respectively. As shown in Fig. 2, the $G$ apparent magnitudes corresponding to the $V$ magnitude limits are 20 and 18 mag for Frolov 1 and NGC 7510, respectively. Thus we adopted these values as cluster photometric completeness limits. During further analyses we used the stars brighter than the completeness $V$ limits adopted for two clusters. 

\begin{figure*}
\centering
\includegraphics[scale=.37, angle=0]{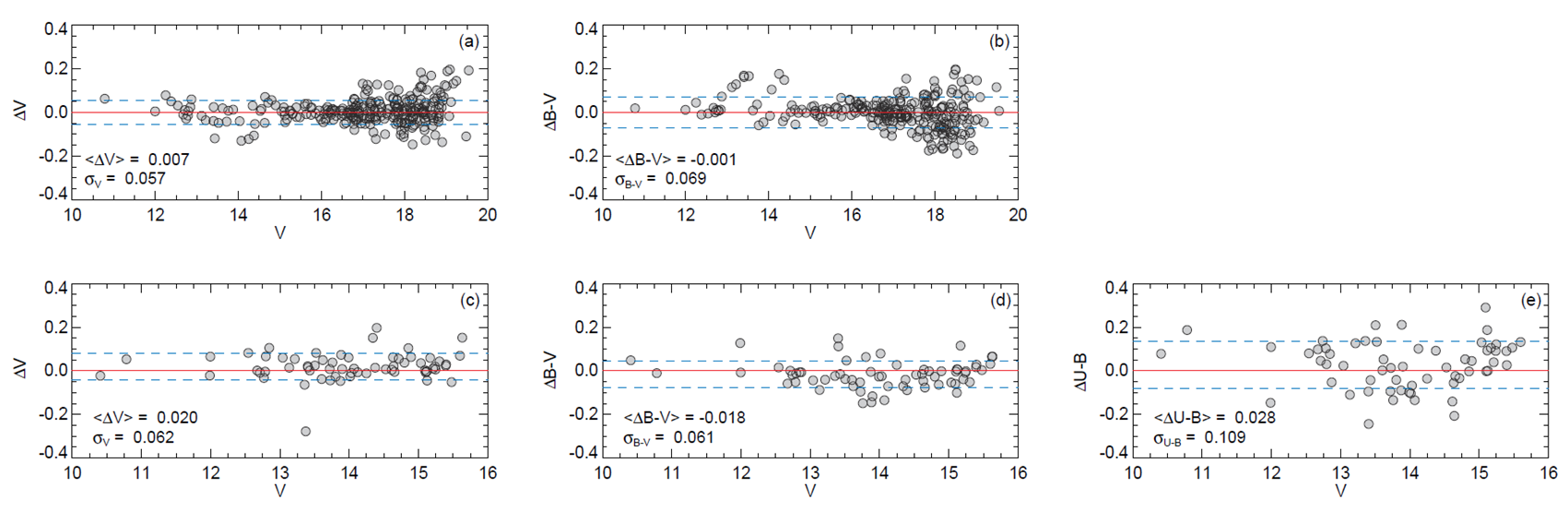}\\
\caption{Comparison of the $V$, $B-V$ and $U-B$ photometric measurements with \citet{Sagar91} (a and b) and \citet{Barbon96} (c, d and e) for NGC 7510.} 
\end {figure*} 

Because Frolov 1 is a poorly studied cluster, there are no available CCD data for this cluster. Therefore we made comparisons to the literature for only our photometric data for NGC 7510 alone, using the CCD {\it BVI} data of \citet{Sagar91} and the photographic {\it UBV} observations given by \citet{Barbon96}. We compared against the $V$ magnitude and $B-V$ colour measurements from these two studies (Fig. 3) and the $U-B$ colour indices with \citet{Barbon96}. Fig 3a-b. denotes the comparison between this study and \citet{Sagar91}. The horizontal axis presents our measurements, while the vertical axis gives the differences between two studies. Based on these comparisons, the mean differences of magnitude and colours are derived as $\langle \Delta V\rangle=0.007\pm 0.057$ and $\langle \Delta {B-V}\rangle=-0.001\pm 0.069$ mag. These measurements applied for only the cross-identified 302 stars brighter than $V=20$ mag in the two studies. The lower panel of Fig. 3c-e represents the comparison between this study and \citet{Barbon96}. The meanings of the horizontal and vertical axes are the same as for Fig. 3a-b. In this comparison, the mean differences in magnitude and colours are derived as $\langle \Delta V\rangle=0.020\pm 0.062$, $\langle \Delta {B-V}\rangle=-0.018\pm 0.061$, and $\langle \Delta {U-B}\rangle=0.028\pm 0.109$ mag. These measurements are based on only the cross-identified 63 stars brighter than $V=16$ mag in both the studies. These comparisons show that photometric measurements in this study are in good agreement with the previous studies.

\subsection{Cluster radius and radial stellar surface density}
The two open clusters analysed in this study have different morphologies: NGC 7510 shows a relatively strong central condensation, while Frolov 1 seems to be a sparse cluster in that its region is dominated by field stars. Even though Frolov 1 has a sparse distribution, structural analyses showed it exhibits the structure of a cluster. In this study, such structural parameters of the two clusters were derived via radial density profile (RDP) fitting to star counts in successive annuli from the cluster centres. We adopted from the SIMBAD database\footnote{http://simbad.u-strasbg.fr/simbad/} central coordinates for the clusters. We constructed circles from a cluster's centre to different angular distances and calculated stellar densities. Then we plotted the angular distance from the centre versus stellar densities, fitting to the RDP model of \citet{King62} through a $\chi^2$ minimisation technique (see Fig. 4). The \citet{King62} formula is expressed as $\rho(r)=f_{bg}+[f_0/(1+(r/r_c)^2)] $. Here, $r$ denotes the centred cluster radius, while $f_0$, $r_c$, and $f_{bg}$ present central density, core radius, and background density respectively. The results of the best fits for Frolov 1 and NGC 7510 are shown in Fig. 4 with dashed blue lines which represent the background density. As seen in Fig. 4 the RDPs are getting flatter at $r=4$ arcmin for Frolov 1 and $r=5$ arcmin for NGC 7510, before merging with the background stellar density. Therefore in further analyses we take into account  only the stars within these radial distances. The RDP results for Frolov 1 are a central stellar density $f_0=4.923 \pm 1.164$, a background stellar density $f_{bg}=4.497\pm0.130$ stars arcmin$^{-2}$, and a core radius $r_c=0.689\pm1.110$  arcmin. Results for NGC 7510 are $f_0=4.771\pm0.022$, $f_{bg}=2.374\pm0.011$ stars arcmin$^{-2}$ and $r_c=2.716\pm0.019$ arcmin. Results are also listed in Table 6.  

\begin{figure}
 \centering
\includegraphics[scale=.7, angle=0]{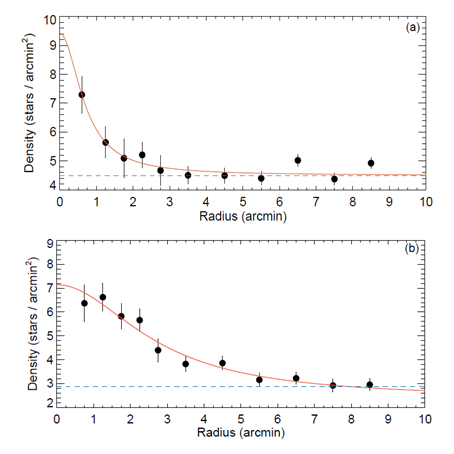}\\
\caption{Stellar density profiles of Frolov 1 (a) and NGC 7510 (b). Errors were determined from $1/\sqrt N$, where $N$ represents the number of stars used in the stellar density estimation. Blue dashed lines denote background stellar density.} \end {figure} 

\begin{table}
\setlength{\tabcolsep}{3pt}
  \centering
  \caption{The structural parameters of the two open clusters according to \citet{King62} model analyses. $f_0$, $r_c$, and $f_{bg}$ are central stellar density, the core radius, and the background stellar densities, respectively.}
    \begin{tabular}{lccc}
\hline
    Cluster & $f_0$ & $r_c$ & $f_{bg}$  \\
            & (stars arcmin$^{-2}$) & (arcmin) & (stars arcmin$^{-2}$) \\
\hline
Frolov 1 & 4.923$\pm$1.164 & 0.689$\pm$1.110 & 4.497$\pm$0.130 \\
NGC 7510 & 4.771$\pm$0.022 & 2.716$\pm$0.019 & 2.374$\pm$0.011 \\
\hline
    \end{tabular}%
\end{table}%

\begin{figure*}[ht]
\centering
\includegraphics[scale=.4, angle=0]{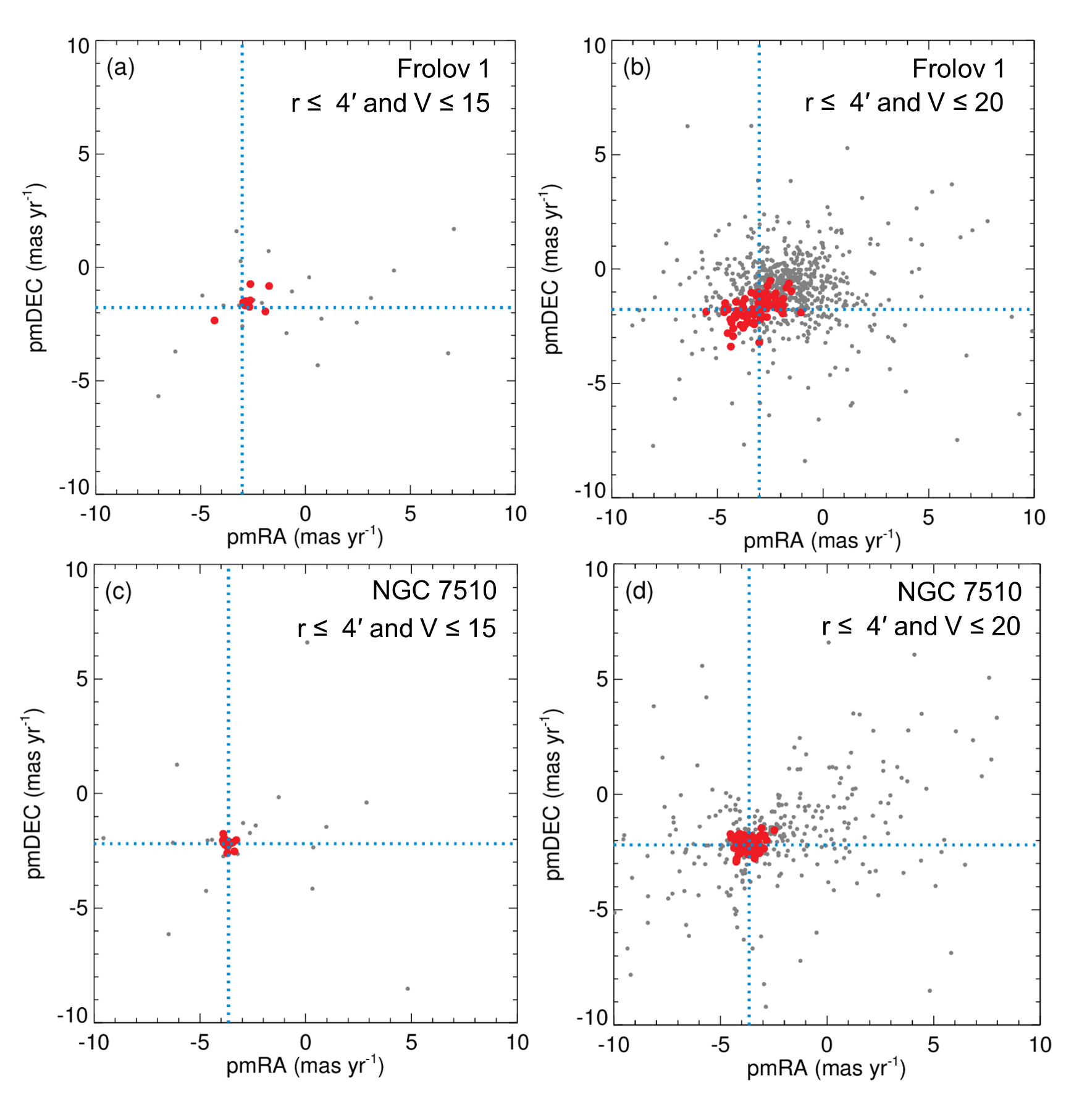}\\
\caption{VPDs of Frolov 1 (a, b) and NGC 7510 (c, d) created as function of $V$ apparent magnitudes and efficient radii. Red dots denote the most probable ($P\geq 0.5$) cluster member stars. The intersection point of blue dotted lines shows the centre of proper motion components of the clusters.} 
\end {figure*}
\newpage

\begin{figure*}
\centering
\includegraphics[scale=.6, angle=0]{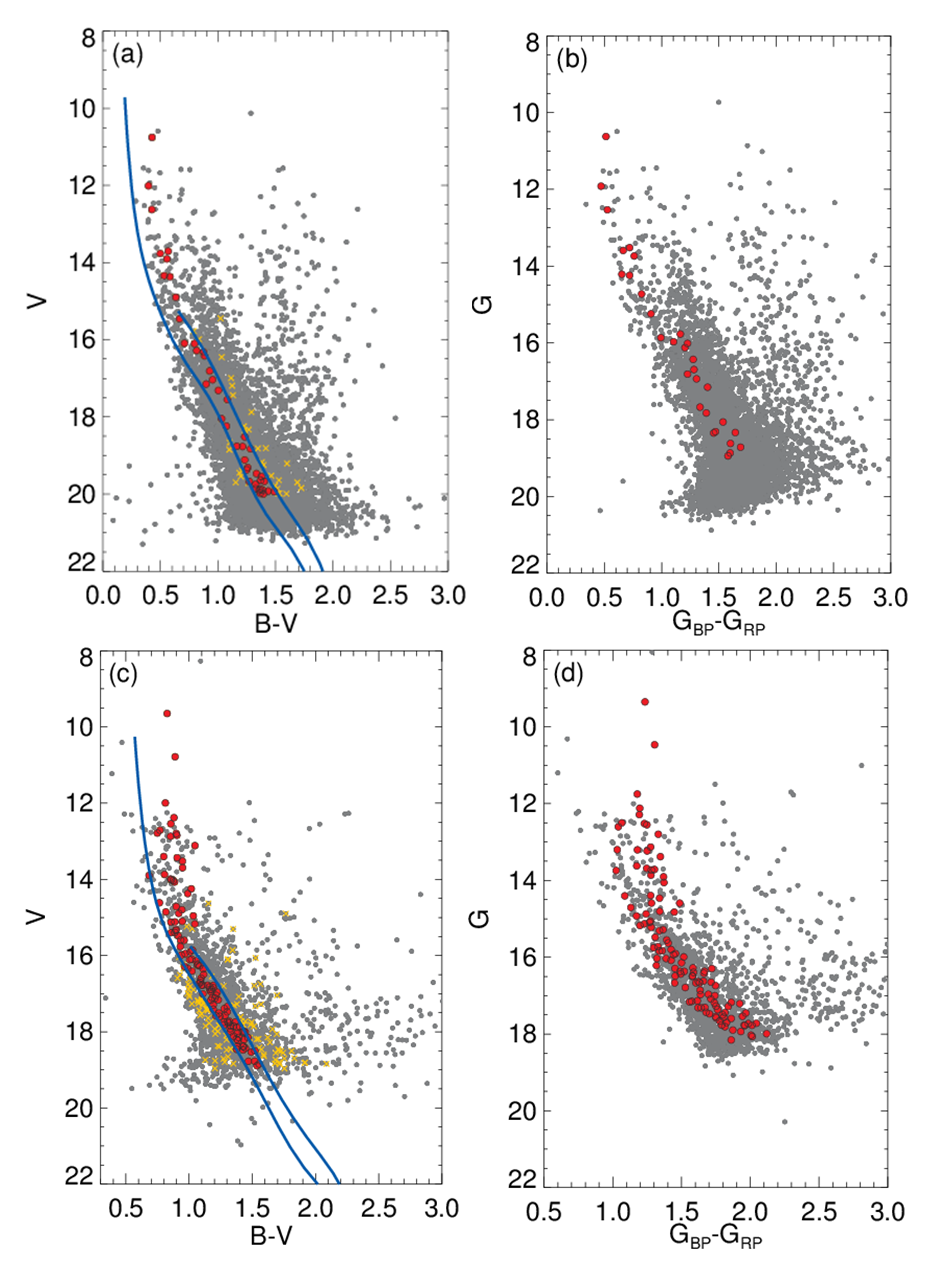}
\caption{$V\times B-V$, $G\times G_{BP}-G_{RP}$ CMDs of Frolov 1 (a, b) and NGC 7510 (c, d). Red dots and yellow crosses indicate the stars with high membership probability ($P\geq0.5$) which lie within $r\leq4$ and  $r\leq5$ arcmin cluster radii determined for Frolov 1 and NGC 7510, respectively. Red dots also show the stars selected considering the binary contamination in the clusters. Grey dots represent less ($P<0.5$) probable cluster stars, as well. In panels (a) and (c), lower and upper blue solid lines represent the ZAMS and  binary sequence, respectively, for two clusters.} 
\end {figure*}

\subsection{Membership Selection}
Contamination by field stars may cause the inaccurate calculation of a cluster's main parameters. For this reason, one should properly separate the cluster members from field stars. In particular, it is important to define the physical members of cluster such as main-sequence stars, giant stars (if they exist in the cluster), and stars that are located near  the main- sequence turn off point before starting astrophysical calculations. The most reliable determination of the physical member stars would be to know the radial velocities, distances, and proper motions of the stars present in cluster field. However, today there is no available sky survey to measure radial velocities of all stars. As cluster stars have the same physical formation conditions they hold similar directions of movement. This makes the proper motion of stars an important tool to determine physical members of a given cluster. Therefore, membership calculations utilising various statistical methods take into account proper motion components as main parameters. In order to calculate the membership probabilities ($P$) in the Frolov 1 and NGC 7510 imaged fields, we used {\it Gaia} DR2 proper motion components of stars and applied the non-parametric method given by \citet{Balaguer98}. This method determines the distributions of field and cluster stars empirically according to their mean cluster and stellar proper motion components. The data distributions were derived considering the technique of kernel estimation with a circular Gaussian kernel function. 

We selected the stars whose membership probabilities are  $P\geq0.5$ as the most probable cluster members and took them into account for further analysis. Before starting the determination of fundamental astrophysical parameters, we constructed vector point diagrams (VPDs) as functions of $V$ apparent magnitudes for the two clusters in Fig. 5 to check the distribution of the cluster members in proper motion space. Generally, the member stars of open clusters are dominant in bright apparent magnitudes, which allows them to be more easily distinguished from field stars. In accordance with this thought, we plotted the VPDs employing stars whose magnitudes are within $V\leq 15$ mag and brighter than the $V$-completeness limits determined above for the two clusters. Considering the examples of the clusters containing bright and faint stars in Fig. 5, it can be seen that their positions in the VPDs do not change and they are clustered in different regions than field stars despite some scattered data.

To take into account the binary star effect in the main-sequence of the clusters, we constructed $V\times B-V$ CMDs of the stars located through Frolov 1 and NGC 7510 before fitting the zero age main-sequence (ZAMS) of \citet{Sung13} to these diagrams of the most probable member stars ($P\geq0.5$). Next we shifted the fitted ZAMS 0.75 mag (Fig. 6) to brighter magnitudes in order to account for binary stars. We made sure that the most likely main-sequence, turn-off, and giant stars of the clusters were selected. The stars whose membership probabilities were $P\geq0.5$, located in the fitted ZAMS regions, and positioned within the clusters' calculated radii described above were adopted as the most likely member stars of the clusters and used in the subsequent analyses. These resulted in 46 and 119 `member' stars for Frolov 1 and NGC 7510, respectively. The $V\times B-V$ CMDs with fitted ZAMS are shown in Fig. 6a,c. Also, we show the distribution of field stars and these most probable member stars on CMDs based on {\it Gaia} DR2 photometry (Fig. 6b,d). Fig. 6 shows that the member stars selected using the astrometric and photometric methods are more scattered in the {\it Gaia} CMDs.  The histograms of the membership probabilities of the stars in the fields of Frolov 1 and NGC 7510 are presented in Fig. 7. The mean proper motion components of stars selected as cluster members in Frolov 1 and NGC 7510 are ($\mu_{\alpha}\cos \delta$, $\mu_{\delta})$=($-3.02\pm 0.10, -1.75\pm 0.08$) and ($\mu_{\alpha}\cos \delta$, $\mu_{\delta}$)=($-3.66\pm 0.07, -2.17\pm 0.06$) mas yr$^{-1}$, respectively.

\begin{figure}
\centering
\includegraphics[scale=0.7, angle=0]{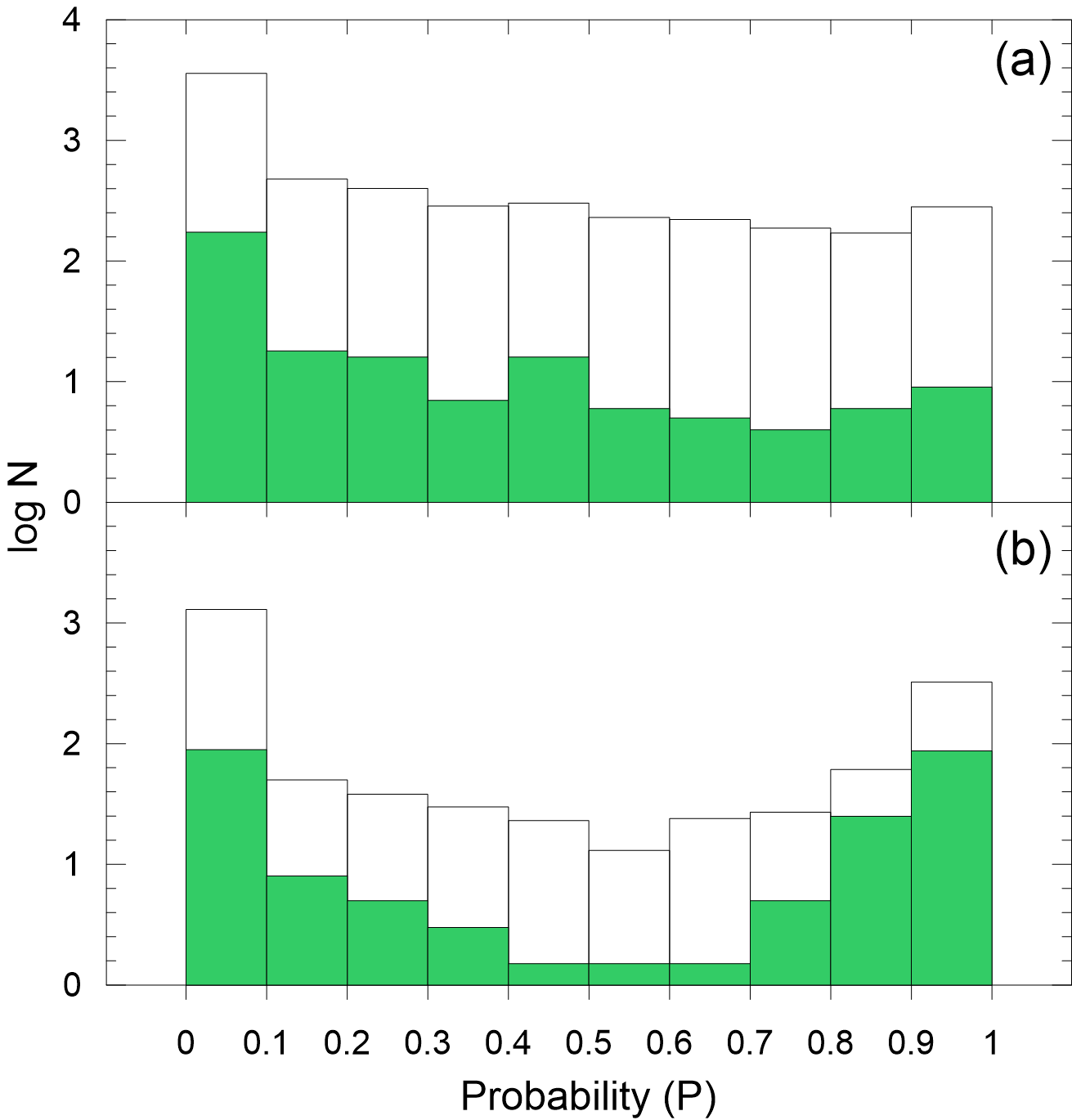}
\caption{The histograms of the membership probabilities of the stars in the fields of Frolov 1 (a) and NGC 7510 (b). The green bars represent the stars that lie within the main-sequence band and efficient cluster radii.} 
\end {figure}

\section{Cluster Parameters}

This section is a summary of the methods utilized to determine the fundamental parameters of Frolov 1 and NGC 7510 \citep[for background details on methodology see][]{Ak16, Bilir16, Bostanci18, Yontan19, Banks20}. We obtained reddening and  metallicity  estimates by fitting observational models on the TCDs, while distance moduli and ages were determined simultaneously via fitting theoretical isochrones on CMDs. 

\begin{figure*}[ht]
\centering
\includegraphics[scale=0.35, angle=0]{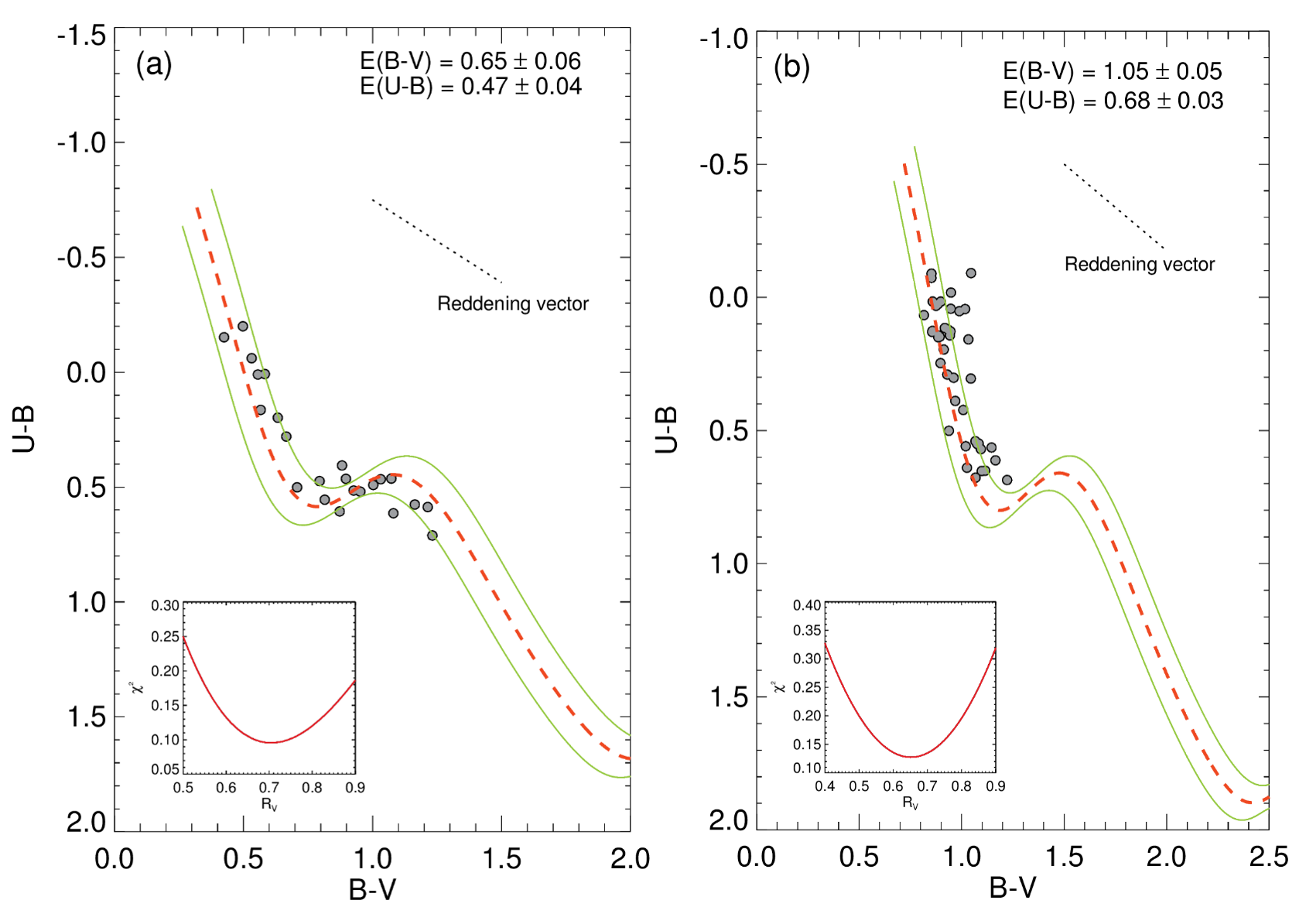}
\caption{$U-B\times B-V$ TCDs constructed for the most probable member main-sequence stars for Frolov 1 (a) and NGC 7510 (b). Red solid lines in the inner panels indicate the $\chi^2$ values corresponding to the $R_V$ results in each cluster. Red dashed lines and green solid lines denote the reddened ZAMS of \citet{Sung13} and $\pm1\sigma$ standard deviations, respectively.} \end {figure*}

\subsection{Reddening}

Because TCDs and CMDs are affected by interstellar reddening, one should first determine $E(U-B)$ and $E(B-V)$ colour excesses. To derive these parameters, we selected $P\geq 0.5$ main-sequence stars within $12.5\leq V\leq 18.5$ and $12\leq V \leq 17$ magnitude ranges for Frolov 1 and NGC 7510, respectively. Then we plotted $U-B\times B-V$ TCDs for these stars and compared their positions with the solar metallicity de-reddened ZAMS of \citet{Sung13}. To obtain the selective absorption coefficient ($R_V=E(U-B)/E(B-V)$) for each cluster, the ZAMS was fitted within $0.4<R_V<0.9$ with steps of 0.01 mag utilizing a $\chi^2$ optimization method (see inner panels of Fig. 8). $R_V$ values corresponding to the minimum $\chi^2$ were accepted for each of the clusters. This method led to the estimated the $R_V$ values of 0.72 and 0.65 for Frolov 1 and NGC 7510, respectively. The fitting process also derived best fitting values for the colour excesses of the two clusters. These colour excesses were determined as $E(B-V)=0.65\pm 0.06$ mag for Frolov 1 and $E(B-V)=1.05\pm 0.05$ mag for NGC 7510. The $E(U-B)$ colour excesses were calculated as $E(B-V) \times R_V$. The presented errors are $\pm 1\sigma$ standard deviations. The TCDs with the best fit ZAMS are shown as Fig. 8.

\begin{figure}
\centering
\includegraphics[scale=.7, angle=0]{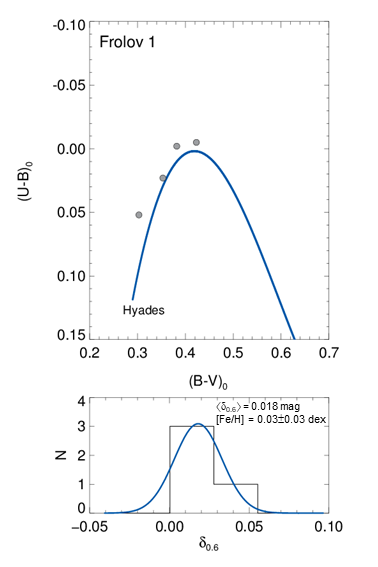}
\caption{The position of four main-sequence stars in Frolov 1 on $(U-B)_0 \times (B-V)_0$ TCD (upper panel) and the histogram for the normalised UV excesses of these stars (lower panel). The solid blue lines in the upper and lower panels denote the main-sequence of Hyades cluster and the Gaussian fit to the histogram, respectively.} 
\end {figure}

\subsection{Metallicity}

Metallicity of the two clusters has not been derived in the earlier studies in the literature. To obtain photometric metallicities we used the method described by \citet{Karaali11}. Because the method utilises F-G type main-sequence stars, we took into account the main-sequence stars whose colour index range is $0.3\leq (B-V)_0\leq0.6$ mag (Eker et al. 2018; 2020) and with membership probabilities $P\geq 0.5$ during these analyses. We constructed de-reddened $(U-B)_0\times(B-V)_0$ TCDs with the selected F-G type stars and the Hyades main-sequence. Then, we calculated the difference between the $(U-B)_0$ colour indices of the cluster stars and Hyades stars whose de-reddened $(B-V)_0$ colour indices are the same. This difference is described as the UV excess of the stars and calculated via the equation $\delta =(U-B)_{0,H}-(U-B)_{0,S}$. Here, H and S represent the Hyades and cluster stars, respectively, which have same $(B-V)_0$ colour indices. Normalising the UV excess differences according to $(B-V)_0 = 0.6$ mag (i.e. $\delta_{0.6}$), we constructed the histogram of normalised $\delta_{0.6}$ values of the stars and then fitted the distribution with a Gaussian. From the Gaussian peak, we determined the metallicity of open clusters using the following equation as given by \citet{Karaali11}:

\begin{eqnarray}
{\rm [Fe/H]}=-14.316(1.919)\delta_{0.6}^2-3.557(0.285)\delta_{0.6}\\ \nonumber
+0.105(0.039).
\end{eqnarray} 

\noindent We could determine photometric metallicity only for Frolov 1 as there are no F-G type main-sequence stars within the colour index $0.3\leq (B-V)_0\leq 0.6$ mag interval for NGC 7510. We identified four F-G type main-sequence stars to calculate the photometric metallicity of Frolov 1. TCD and normalised $\delta_{0.6}$ UV-excesses distribution are shown in Fig. 9. Uncertainty of the measurements is the statistical uncertainty $\pm 1\sigma$ width of the best fit Gaussian model. The metallicity [Fe/H] which represents the peak value for the $\delta_{0.6}$ distribution was determined as [Fe/H]=$ 0.03 \pm 0.03$ dex for Frolov 1. Going forward, in the absence of additional information, we adopted the metallicity of NGC 7510 as the solar value ([Fe/H] = 0 dex).

\begin{figure*}
\centering
\includegraphics[scale=.49, angle=0]{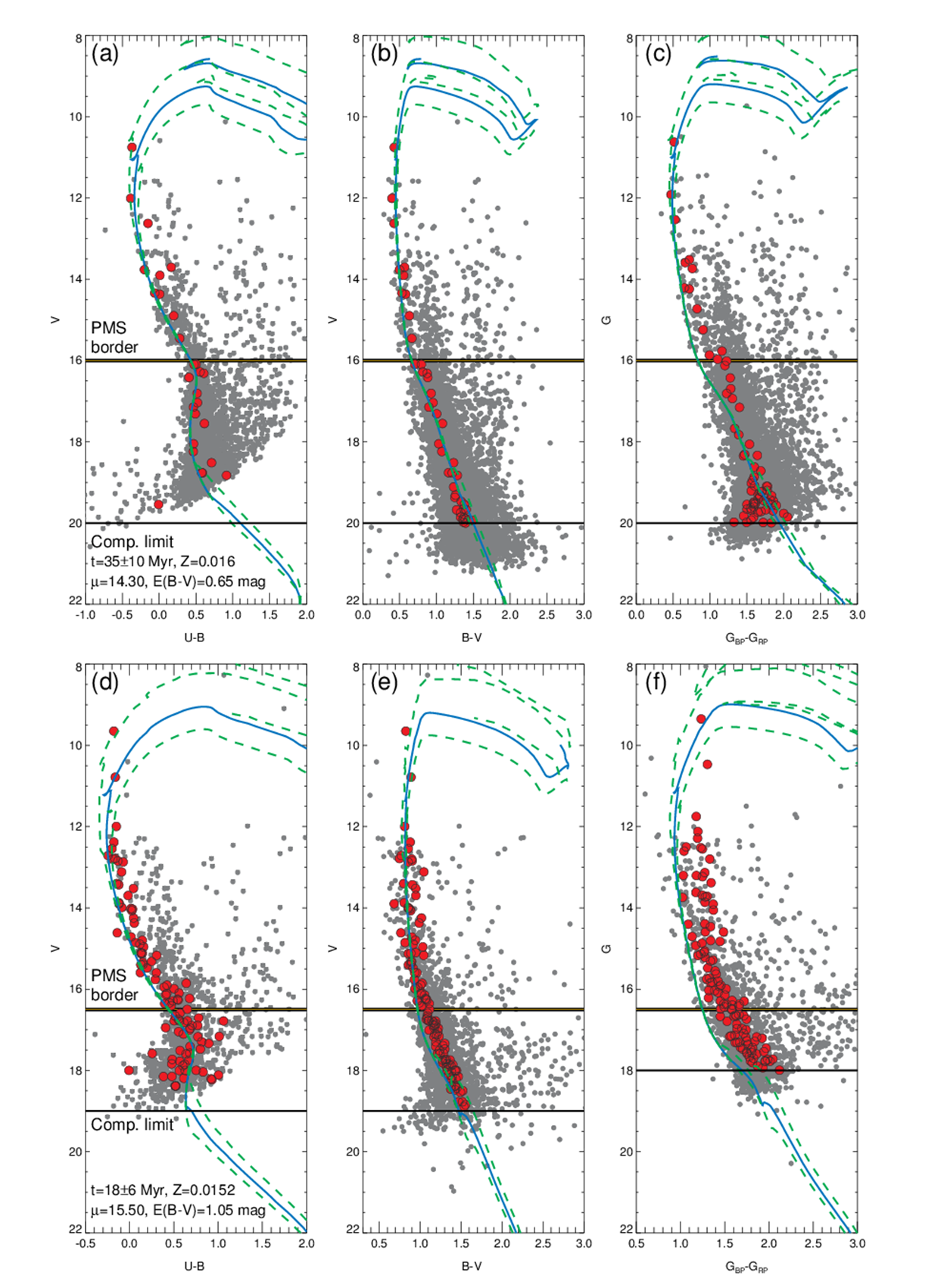}
\caption{CMDs for the stars through the field of Frolov 1 (a, b and c) and NGC 7510 (c, d and e). Red dots indicate the selected member stars of clusters, while grey dots represent less probable cluster members or field stars. Blue solid and green dashed lines indicate the best fit isochrones of ages and errors, respectively. Thick and thin black horizontal lines are the borders of PMS and completeness limits, respectively.} 
\end {figure*}

To derive the age of the clusters, we transformed the [Fe/H] metallicities to the mass fraction $Z$, considering the following equations that analytically improved PARSEC isochrones (as given by Bovy\footnote{https://github.com/jobovy/isodist/blob/master/isodist/\\Isochrone.py}):

\begin{equation}
Z_X={10^{{\rm [Fe/H]}+\log \left(\frac{Z_{\odot}}{1-0.248-2.78\times Z_{\odot}}\right)}},
\end{equation}      

\begin{equation}
Z=\frac{(Z_X-0.2485\times Z_X)}{(2.78\times Z_X+1)}.
\end{equation} 
Here, $Z$ and $Z_X$ represent all elements heavier than helium and the intermediate operation function, respectively. $Z_{\odot}$ denotes the solar mass fraction, which was taken as 0.0152 \citep{Bressan12}. As a result of these calculations, we measured $Z=0.016$ for Frolov 1.

\subsection{Distance moduli and the ages of the clusters}

Distance moduli and the ages of the clusters were determined by comparing the $V\times U-B$, $V\times B-V$, and $G\times G_{BP}-G_{RP}$ CMDs with theoretical isochrones procured from the PARSEC synthetic stellar library \citep{Bressan12}. For Frolov~1, since the reddening and metallicity were derived separately from TCDs and UV excesses according to distribution of member stars on these diagrams, as mentioned above, we kept reddening and metallicities as constants during the distance moduli and age calculations. To avoid degeneracies between the parameters it is important to measure astrophysical parameters separately. Indeed reddening and age values in particular can suffer from degeneracies and uncertainties of these parameters can be higher than expected \citep{Bilir10, Yontan15, Bostanci15}. In order to achieve the best fit of isochrones during the calculations, we fitted them on CMDs with an emphasis on the main-sequence and turn-off members of the clusters. Errors estimates for the ages were determined visually by shifting isochrones to cover all main-sequence member stars. We used the relations given by \citet{Carraro17} in determining the error estimates for the distance moduli and distances. For the age estimates made using the $G\times G_{BP}-G_{RP}$ CMD, the PARSEC isochrone has been reddened according to the equation of $E(G_{BP}-G_{RP})= 1.2803\times E(B-V)$. We calculated the coefficient (1.2803) in the equation from \citet{Wang19}'s selective absorption coefficients.The best fitting ischrone for Frolov 1 was $Z = 0.016$, giving an age of $t=35 \pm 10$ Myr and a distance of $d=2,864 \pm 254$ pc. For NGC 7510, the best fitting isochrone was $Z=0.0152$, giving an age of $t=18 \pm 6$ Myr and a distance of $d=2,818 \pm 247$ pc. These isochrones are over-plotted in the $V\times U-B$, $V\times B-V$ and $G\times G_{BP}-G_{RP}$ CMDs given as Fig. 10. Different distance estimations will be discussed later in connection with the {\it Gaia} measurements in Section 4.4.

Since Frolov 1 and NGC 7510 are relatively young ($t<40$ Myr) open clusters, there can still be cluster members that are pre-main sequence stars (PMS). This may affect the determination of the astrophysical parameters of the clusters. In this study we considered the evolutionary properties of the PARSEC isochrones which were fitted to member stars in different luminosity classes, leading to possible PMS limits for the two clusters being determined. These limits are $V=16$ and $V=16.5$ mag for Frolov 1 and NGC 7510, respectively, and are shown as thick horizontal lines in Fig. 10. Additionally the $V$ and $G$ completeness magnitude limits of the two clusters are shown as the thin horizontal black lines on the same figures. Since {\it Gaia} $G_{BP}$ and $G_{RP}$ bands are located in blue and red part of the optical electromagnetic spectrum and they are wider than the {\it UBV} bands, pre-main sequence stars are more prominent on {\it Gaia} CMDs. In estimating the age and distance of the two clusters, it was concluded that the PARSEC isochrones that had been fitted to the most probable cluster member stars were not affected by the PMS stars within the theoretical limits in this study (see Fig. 10).  

\subsection{Distance comparisons and mean proper motions of the clusters}

We calculated mean values of the proper motion components and {\it Gaia} distances for the two clusters from {\it Gaia} DR2 astrometric data \citep{Gaia18}. Using the linear equation of $d({\rm pc})=1000/\varpi$ (mas), we transformed the {\it Gaia} trigonometric parallaxes to {\it Gaia} distances for each star that was selected as a member of Frolov 1 and NGC 7510. Then we constructed the distance histograms, fitted Gaussian curves to them (Fig. 11), and derived mean {\it Gaia} distances ($d_{\rm Gaia}$). The stated uncertainties in distances are one standard deviation. 

Just after the {\it Gaia} DR2 \citep{Gaia18}, many researchers found different trigonometric parallax zero-point values based on different objects \citep[i.e.][]{Arenou18, Lindegren18, Riess18}. These values vary between $-0.029$ and $-0.082$ mas \citep{Lindegren18, Stassun18}. \citet{Lindegren18} used a sample of quasars with {\it Gaia} DR2 and proposed a $-0.029$ mas value for the offset correction to trigonometric parallaxes. However, \citet{Stassun18} used a {\it Gaia} DR2 data of eclipsing binary systems whose magnitudes are brighter than $G\leq 12$ mag and with distances within $0.03<d<3$ kpc, finding evidence for a systematic offset of $-0.082$ mas. The zero point-offset corrections of {\it Gaia} DR2 parallaxes can be tested by considering the distance of the open clusters determined by isochrone fitting method and the distances obtained from {\it Gaia} DR2 parallaxes. In this study, to examine the offsets of the two cluster distances, we converted the distances calculated by isochrone fitting and {\it Gaia} DR2 trigonometric parallaxes (Table 7) to parallaxes using the equation of $\varpi ({\rm mas})=1000/d$ (pc) and the differences between them ($\varpi_{\rm iso}-\varpi_{\rm Gaia}$) were calculated. For Frolov 1, isochrone fitting and {\it Gaia} distances translate to $\varpi_{iso}=0.349\pm 0.031$ and $\varpi_{\rm Gaia}=0.357\pm 0.017$ mas, respectively. These values are $\varpi_{iso}=0.355\pm 0.031$ and $\varpi_{\rm Gaia}=0.290\pm 0.041$ mas for NGC 7510. The differences in the sense are $\varpi_{\rm iso}-\varpi_{\rm Gaia}=-0.008$ and $\varpi_{\rm iso}-\varpi_{\rm Gaia}=0.065$ mas for Frolov 1 and NGC 7510, respectively. Both these results are in good agreement with the values given in different studies \citep[i.e.][]{Arenou18, Lindegren18, Riess18, Stassun18}. As a result of these distance comparisons we concluded that trigonometric parallax corrections are not statistically notable. In this study, we did not apply a zero-point offset correction for {\it Gaia} DR2 trigonometric parallaxes of the cluster member stars. \citet{Arenou18} suggest that examining different samples of sources can change the derived parallax zero-point values. 

\begin{figure}[ht]
\centering
\includegraphics[scale=.5, angle=0]{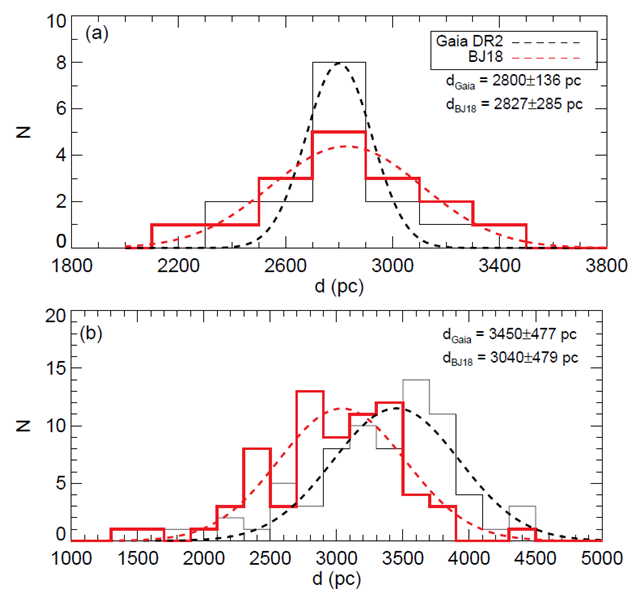}
\caption{The distance distribution of the cluster member stars for Frolov 1 (a) and NGC 7510 (b) for the methods of linear \citep[Gaia DR2,][]{Gaia18} and geometric distances \citep[BJ18,][]{Bailer-Jones18}. The dashed lines indicate the best Gaussian fits to the histograms.} 
\end {figure}

Recently \citet{Bailer-Jones18} (hereafter BJ18) indicated that using a linear method for measuring distances from trigonometric parallaxes does not give precise results. Based on this finding, they improved the geometric approach to calculate distances of stars, making use of probability distributions for the distances of stars. \citet{Bailer-Jones18} applied a geometric approach to 1.33 billion objects in the {\it Gaia} DR2 catalogue) to estimate distances and distance errors. In this context, to understand how different results from linear and geometric approaches are, we sourced distances of cluster member stars given in the catalogue of \citet{Bailer-Jones18}. In calculating the mean distances of the two clusters, {\it Gaia} stars with a relative parallax error of less than 0.5 were taken into account. Then we constructed the two distance  histograms of these values and fitted Gaussian curves to derive the mean distances. We compared the two distance distributions, as shown in Fig. 11. Calculated distances from three different methods are listed in Table 7 for the two clusters. In Table 7, we can clearly see that the distances determined with isochrone fitting are more compatible with the ones from \citet{Bailer-Jones18}. 

In this study, as mentioned in Section 3.3, we used the methods given by \citet{Balaguer98} and calculated membership probabilities based on {\it Gaia} DR2 proper motions. \citet{Cantat-Gaudin18} studied 401,488 stars in 1,229 open clusters and determined both their membership probabilities and mean proper motions from photometric and astrometric data of the {\it Gaia} DR2 \citep{Gaia18}. To estimate membership probabilities, they applied a different calculation method which is given by \citet{Krone-Martins14}. As Frolov 1 is not examined in the study of \citet{Cantat-Gaudin18}, we could only compare the results of NGC 7510 with their results. In Table 7, it can be clearly seen that the mean proper motion values calculated in this study and \citet{Cantat-Gaudin18} are in good agreement. However, methods differ from each other across the three studies, so the coherence of the results indicates that the member stars were selected accurately and isochrone fitting is still reliable for distance estimations of open clusters. Also, one should remember that the astrophysical parameters determined with independent methods in this study, as well as distances calculated from different three methods, are compatible. So, we can conclude that the results obtained in the study have not suffered from parameter degeneracy.

\subsection{Mass functions}

We used the best fitting PARSEC isochrones for the two clusters to determine the present-day mass functions (MFs). Taking into account main-sequence stars, we first specified a high degree polynomial function between theoretical $V$ band absolute magnitudes and masses, and identified an absolute magnitude-mass relation. Then, we calculated $V$ band absolute magnitudes of cluster member stars according to the derived distance moduli of the clusters. Following this, using the derived relation we measured masses of the most likely cluster member stars for each cluster. Taking the relation of $\log(dN/dM)=-(1+X)\times \log M + {\rm constant}$ (where, $dN$ and $X$ are number of stars in a mass interval $dM$ with central mass $M$ and slope of the MF, respectively) we determined mass function slopes. MF slopes were derived within $1\leq M/ M_{\odot}\leq 5$ as $X=-1.21\pm0.18$ for Frolov 1, and across $2\leq  M/ M_{\odot}\leq 10$ $X=-1.42\pm 0.27$ for NGC 7510. The resulting MF slopes are in good agreement with \citet{Salpeter55}'s value of $-1.35$. The best linear fits are shown as Fig. 12 and results are listed in Table 8 for both clusters.

\begin{figure}[ht]
\centering
\includegraphics[scale=.5, angle=0]{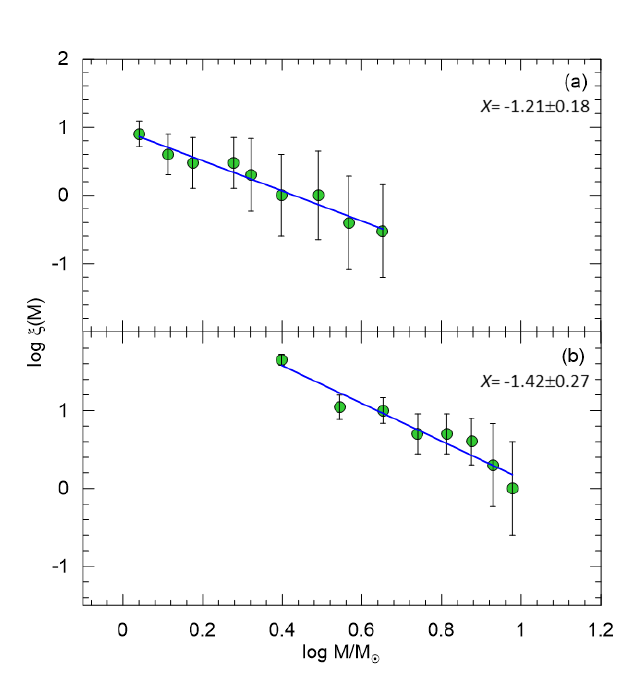}
\caption{Mass functions of Frolov 1 (a) and NGC 7510 (b) derived from the most probable cluster member stars. Solid blue lines are the mass functions of the two clusters.} 
\end {figure}

\begin{table*}[t]
\setlength{\tabcolsep}{4.5pt}
\renewcommand{\arraystretch}{1.75}
  \centering
\scriptsize{
  \caption{Calculated distances and proper motion results for different methods for the two clusters.}
\begin{tabular}{l|cccc|ccc|c}
\hline
                  & \multicolumn{4}{c}{This Study} &  \multicolumn{3}{c}{\citet{Cantat-Gaudin18}} & \citet{Bailer-Jones18} \\
\hline
     Cluster      & $d_{Iso}$ & $d_{Gaia}$ & $\mu_{\alpha}\cos \delta$ & $\mu_{\delta}$ & $d$ & $\mu_{\alpha}\cos \delta$ & $\mu_{\delta}$ & $d_{BJ18}$ \\
                  &    (pc)   & (pc) & (mas~yr$^{-1}$)  & (mas~yr$^{-1}$) & (pc) & (mas~yr$^{-1}$) & (mas~yr$^{-1}$) & (pc) \\
\hline
    Frolov 1       &  2,864$\pm$254   &  2,800$\pm$136  & $-$3.02$\pm$0.10 & $-$1.75$\pm$0.08 &    --   & -- & -- &  2,827$\pm$285 \\
    NGC 7510 &  2,818$\pm$247 & 3,450$\pm$477  & $-$3.66$\pm$0.07 & $-$2.17$\pm$0.06 &  3,178$^{+1,480}_{-767}$ & $-$3.66$\pm$0.13 & $-$2.19$\pm$0.13 & 3,040$\pm$479 \\
\hline
    \end{tabular}%
}
  \label{tab:addlabel}%
\end{table*}%

\begin{table}
\setlength{\tabcolsep}{7pt}
\renewcommand{\arraystretch}{1.5}
  \centering
  \caption{The slopes ($X$) of the mass functions of Frolov 1 and NGC 7510. $N$ is number of stars used in the calculations.}
    \begin{tabular}{lccc}
\hline
    Cluster & $N$ & $X$ & Mass Range \\
\hline
    Frolov 1 &  24 & -1.21$\pm$0.18 & 1 $<M/M_{\odot}<$ 5 \\
    NGC 7510 &  83 & -1.42$\pm$0.27 & 2 $<M/M_{\odot}<$ 10 \\
\hline
    \end{tabular}
\end{table} 


\section{Summary and Conclusion}

We performed a study based on CCD {\it UBV} photometric and {\it Gaia} DR2 photometric and astrometric data of the open clusters Frolov 1 and NGC 7510. In Table 1 we listed derived basic parameters. We summarise the results of analyses as follows:

\begin{enumerate}[1.]
\item We determined estimates for the reddening, metallicity, distance modulus, distance, and age of each cluster through the analysis of  CCD {\it UBV} photometric data.

\item Since the determination of cluster parameters is affected by many conditions, such as dense stellar regions, membership selection, high/differential extinction, and different analysis methods, degeneracies can result between reddening and age parameters \citep{Anders04, King05} and in turn cause large discrepancies in estimates for the same cluster as studied by different authors. To avoid parameter degeneracy, we analysed the two clusters using independent methods and obtained the main astrophysical parameters separately.

\item Fitting the RDP of \citet{King62}, we derived structural properties for each cluster (Table 6). During the analyses we used the radii where stellar densities are flattened and converge with the background stellar density as the outer cutoff limits. These values are $r=4$ and $r=5$ arcmin for Frolov 1 and NGC 7510, respectively. Although the central stellar density and background stellar density values of Frolov 1 show that the cluster has a low stellar density concentration, from the RDP fitting results we conclude that Frolov 1 shows cluster structure. 

\item In order to obtain physical member stars of the clusters, we used {\it Gaia} DR2 proper motion values. In addition to this, we derived mean proper motion components for Frolov 1 as ($\mu_{\alpha}\cos \delta, \mu_{\delta}) = (-3.02\pm 0.10, -1.75\pm 0.08$) mas yr$^{-1}$, and for NGC 7510 as $(\mu_{\alpha}\cos \delta, \mu_{\delta}) = (-3.66\pm 0.07, -2.17\pm 0.06$) mas yr$^{-1}$. We only could compare the results of NGC 7510 with \citet{Cantat-Gaudin18} and found good agreement between the two studies. 

\item From the comparison of TCDs of the two clusters with the ZAMS \citep{Sung13}, we derived colour excesses $E(B-V)= 0.65 \pm 0.06$ mag for Frolov 1 and $E(B-V)=1.05 \pm 0.05$ mag for NGC 7510.

\item Photometric metallicity could be obtained for Frolov 1 as [Fe/H]=$ 0.03 \pm 0.03$ dex, which corresponds to $Z=0.016$. Since there is a lack of F-G stars within $0.3<(B-V)_{\rm 0}<0.6$ mag range in NGC 7510, we adopted a metallicity for the cluster as being the solar value ([Fe/H]=0 dex, corresponding to $Z\simeq 0.015$).

\item Taking reddening and metallicity parameters as constants, we derived distance moduli and ages of two clusters via fitting theoretical isochrones to $V\times U-B$, $V\times B-V$, and $G\times G_{BP}-G_{RP}$ CMDs. We found the distance modulus for Frolov 1 as being  $\mu_V=14.30 \pm 0.20$ mag ($d=2,864 \pm 254$ pc) together with an age $t=35 \pm 10$ Myr, while for NGC 7510 we found $\mu_V=15.50\pm 0.20$ mag ($d=2,818 \pm 247 $ pc) and $t=18 \pm 6$ Myr respectively. These results are in agreement with those given by different authors.

\item Mass functions were derived as $X=-1.21\pm 0.18$ for Frolov 1 and $X=-1.42 \pm 0.27$ for NGC 7510. These mass function slopes are in good agreement with the value of \citet{Salpeter55}. 

\end{enumerate}

\section{Acknowledgments}
This study has been supported in part by the Scientific and Technological Research Council (T\"UB\.ITAK) 119F014. We thank T\"UB\.ITAK for partial support towards using the T100 telescope via project numbers 15AT100-738 and 18CT100-1396. We also thank the on-duty observers and members of the technical staff at the T\"UB\.ITAK National Observatory (TUG) for their support before and during the observations. This research made use of the WEBDA database, operated at the Department of Theoretical Physics and Astrophysics of the Masaryk University. This work also made use of data from the European Space Agency (ESA) mission {\it Gaia}\footnote{https://www.cosmos.esa.int/gaia}, processed by the {\it Gaia} Data Processing and Analysis Consortium  (DPAC)\footnote{https://www.cosmos.esa.int/web/gaia/dpac/consortium}. Funding for the DPAC has been provided by national institutions, in particular those participating in the {\it Gaia} Multilateral Agreement.


\end{document}